\documentclass[ draft
  ]
  {aipproc}

\layoutstyle{6x9}

\usepackage[fleqn]{amsmath}
\usepackage{bbm}

\newcommand{\E}[1]{\ensuremath{\text{E}_{#1}}} 
\newcommand{\G}[1]{\ensuremath{\text{G}_{#1}}}
\newcommand{\SO}[1]{\ensuremath{\text{SO}\!\left(#1\right)}}
\newcommand{\SU}[1]{\ensuremath{\text{SU}\!\left(#1\right)}}
\newcommand{\U}[1]{\ensuremath{\text{U}\!\left(#1\right)}}
\newcommand{\Z}[1]{\ensuremath{\mathbbm{Z}_{#1}}} 

\begin{document}
{\normalsize \hspace{-0.55cm}
DESY-07-136\hfill\mbox{}\\
August 2007\hfill\mbox{}\\}
\vspace{-1cm}
\title{Gauge-Higgs Unification from the Heterotic String}


\author{Jonas Schmidt}{
  address={Deutsches Elektronen-Synchrotron DESY, Notkestrasse 85, 22603 Hamburg, Germany} }



\begin{abstract}
We present a 6D orbifold model on $T^2/\Z2$ which emerges as intermediate step in the compactification of the
heterotic string to the supersymmetric standard model in four dimensions.
It has $\SU6$ gauge symmetry in the bulk and two pairs of inequivalent fixed points with unbroken $\SU5$ and
$\SU2 \times \SU4$ symmetry, respectively. All anomalies are cancelled by the Green-Schwarz mechanism. Two 
quark-lepton generations are located at the $\SU5$ branes, the third family is composed of split bulk hypermultiplets.
The model has vacua with partial or full gauge-Higgs unification and non-vanishing Yukawa couplings which
generically avoid the unsuccessful $\SU5$ mass relations.
\end{abstract}

\maketitle


\section{Introduction}
A completion of the standard model in terms of string theory may reveal an intermediate stage of grand unification. This idea is supported by
the symmetries and the matter content of the standard model as well as the unification of gauge couplings at the scale $M_{\rm GUT} \simeq 2 \times 10^{16}$
GeV for the minimal supersymmetric standard model (MSSM).
Here we present the construction of such an intermediate grand unifying theory (GUT) in six dimensions~\cite{bls07} and discuss 
some interesting aspects of its
local GUT structure.

The model is based on a $\Z{\rm 6-II}$ orbifold compactification of the $\E8 \times \E8$
heterotic string which leads to the MSSM in four dimensions \cite{bhx06-1,bhx06-2}. 
Four extra dimensions are compactified on $T^4/\Z3$, where $T^4$ is the Lie algebra lattice of $\G2 \times \SU3$
 and the $\SU3$ torus is supplemented by a Wilson line.
This leads to a six-dimensional (6D) supergravity theory with unbroken gauge group
\begin{equation}
{\rm G_6}~=~\SU6\times\U1^3\times [\SU3\times\SO8\times\U1^2]\;,
\end{equation}
where the brackets denote the subgroup of the second $\E8$. Additionally to  gravity, dilaton and vector multiplets it has
$\mathcal{N}=2$ hypermultiplets\footnote{All $\U1$ charges are surpressed in the following. The full list is given in \cite{bls07}.}
\begin{equation}
\left({\bf 20};1,1\right) +
\left(1;1,{\bf 8}\right) +
\left(1;1,{\bf 8}_s\right) + \left(1;1,{\bf 8}_c\right) +
4\times \left(1;1,1\right) 
\label{eq:spec-bulk-1}
\end{equation}
from the untwisted sector. Furthermore, there are three fixed points in the $\SU3$ plane. 
They all have unbroken $\SO{14} \times \U1 \times [\SO{14} \times \U1]$ symmetry, differently embedded into $\E8 \times \E8$. Each fixed point contributes three copies of
\begin{subequations}
\begin{align}
({\bf 14},1;1,1) &= ({\bf 6};1,1) + ({\bf \bar{6}};1,1)
+ 2\times (1;1,1)\;, \\
(1,1;{\bf 14},1) &= (1;{\bf 3},1) + (1;{\bf \bar{3}},1)
+ (1;1,{\bf \hat{8}})
\end{align}
\label{eq:spec-bulk-2}
\end{subequations}
to the 6D theory. At the three $\SU3$ fixed points, $(1;1,{\bf \hat{8}})$ corresponds to
$(1;1,{\bf 8})$, $(1;1,{\bf 8}_s)$ and $(1;1,{\bf 8}_c)$,
respectively. Finally, there are oscillator states for the two small
compact planes which yield two non-Abelian singlet hypermultiplets for each fixed point.
\begin{figure}[t]
\includegraphics[height=7cm]{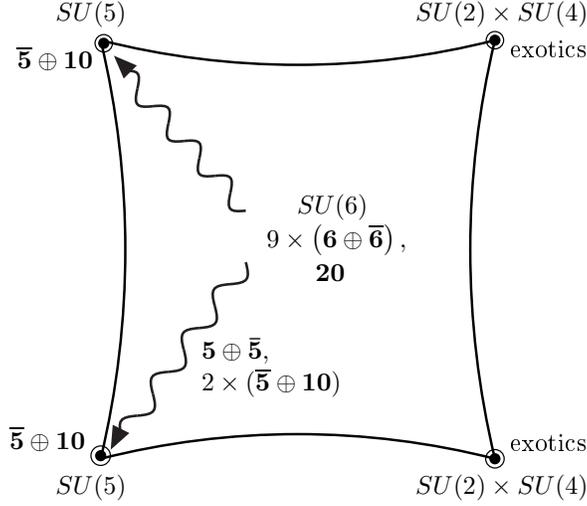}
\caption{The orbifold $T^2/\Z2$. The arrows indicate the local reduction of bulk states at the two equivalent $\SU5$-fixed points ($n_2=0$)
to one pair of Higgs fields (${\bf 5} \oplus {\bf \bar5}$) and two bulk quark-lepton families (${\bf \bar5} \oplus {\bf 10}$),
due to chiral orbifold projections and decoupling. All exotics located at the remaining two fixed points ($n_2=1$) can be decoupled. 
\label{fig:orbi}}
\end{figure}
Compactification from six to four dimensions occurs on the orbifold $T^2/\Z2$, where $T^2$ is the Lie algebra
lattice of $\SO4$ (cf. Figure~\ref{fig:orbi}).
Due to the presence of one Wilson line in this torus there are two inequivalent pairs of fixed points,
labelled by $n_2=0, 1$, with local gauge groups
\begin{align}
 n_2=0: \; \; & G_0 = \SU5 \times \U1^4 \times [ \SU3 \times \SO8 \times \U1^2] \; ,\\
 n_2=1: \; \; & G_1 = \SU2 \times \SU4 \times \U1^4 \times [ \SU2' \times \SU4' \times \U1^4] \; .
\end{align}
The four dimensional gauge group is given by the intersection of the local symmetries. 
At the fixed points we find localized states which transform under the non-Abelian symmetries as 
\begin{align}
	 n_2=0: \;  & ({\bf \bar5};1,1) + ({\bf 10};1,1) \nonumber
	 	  + (1;1,{\bf 8_c}) + 2 \times (1;{\bf 3},1) + 2 \times (1;{\bf \bar3},1) + 9 \times (1;1,1) ,\\
	n_2=1: \;  & 4 \times ({\bf 2},1;1,1) + 16 \times (1,1;1,1) \nonumber \;.
 \end{align}
 All exotics are localized at $n_2=1$ and can be decoupled from the zero mode spectrum. Phenomenologically interesting
 vacua and the four dimensional limit of the model were studied in \cite{bhx06-2}. 
 Here we focus on the anisotropic limit of the compactification in which the $\SO4$ torus is considerably larger than the other two tori.
 This introduces an intermediate field theoretical orbifold GUT in six dimensions. Finally, at string scale energies, all ten dimensions can be
 resolved and full heterotic string theory has to be considered.
 
\section{Anomalies}
The cancellation of anomalies in string derived models follows by application of the Green-Schwarz (GS) mechanism~\cite{gs84}. 
For the bulk and brane anomalies of our 6D heterotic orbifold model~\cite{e94,lnz04,abc03} this relies on
gauge and Lorentz variations of the bulk two-form $B_2$. 
All contributions from box diagrams in the bulk and local triangle diagrams at the fixed points add to reducible anomaly 
polynomials
\begin{align}
	I_8^{\rm bulk} &\sim X_4 \wedge Y_4 \;, & I_6^f &\sim X_4^f \wedge Y_2^f \;.
	\label{eq:i8i6}
\end{align}
Here a subscript $n$ denotes a (formal) $n$-form and $f$ a projection onto the fixed point $f$. Note that the factorizations~(\ref{eq:i8i6})
invoke $\sim 500$ conditions in the case of our model which imposes a highly non-trivial
check for the spectrum presented before. Furthermore, the explicit calculation of the two-forms $Y_2^f$ allow the identification of local
anomalous $\U1$'s at the fixed points. They lead to the following coefficients $\xi_{n_2}$ of local Fayet-Iliopoulos (FI) $D$-terms:
\begin{align}
	\xi_0 & = 2 \, \frac{g M_{\rm P}^2}{384 \pi^2} \; , & \xi_1 & = \frac{g M_{\rm P}^2}{384 \pi^2} \; .
\end{align}
It is known that localized FI terms lead to an instability of bulk fields and spontaneous localization towards the fixed points~\cite{lnz04}.
It is intriguing that the mass scale of the FI terms is of
the order of the grand unification scale, $M_{\rm P}/\sqrt{384 \pi^2} \sim M_{\rm GUT}$.

\section{Local $\SU5$ Unification}
All bulk multiplets~(\ref{eq:spec-bulk-1}),~(\ref{eq:spec-bulk-2}) are $\mathcal{N}=2$ hypermultiplets $H=(H_L,H_R)$, where $H_L$ and $H_R$ are
$\mathcal{N}=1$ left and right chiral multiplets, respectively. 
The multiplets (\ref{eq:spec-bulk-2}) come in three copies, due to three fixed points in the $\G2$ plane. Thus 
even after the chiral projection
onto $n_2=0$ it is still possible to form singlets $H_L H_R^c$ . A large number of fields 
can in this way easily be decoupled by giving a vacuum expectation value (VEV) to only three singlet fields,
two of which are localized at $n_2=0$. This yields localized $\mathcal{N}=1$ chiral multiplets $3 \times {\bf 5} \oplus 5 \times {\bf \bar5} \oplus 2 \times {\bf 10}$ from the bulk,
including a ${\bf 5} \oplus {\bf \bar5}$ pair originating from the $\SU6$ gauge multiplet. 
In a second step, one
can again form two singlets ${\bf 5} \, {\bf \bar5}$ and specify VEVs which induce masses for the involved fields. 
Thus finally one ${\bf 5} \oplus {\bf \bar5}$ pair of Higgs fields and two $\SU5$ families ${\bf \bar5} \oplus {\bf 10}$ remain light
after decoupling (cf.~Figure~\ref{fig:orbi}).

Depending on the origin of the Higgs multiplets the model can have full, partial or no gauge-Higgs unification. As an example, we
realized partial gauge-Higgs
unification and the decoupling of all exotic doublets by giving VEVs to 19 singlet fields. To $\mathcal{O}(8)$ in the fields 
the Yukawa couplings of the zero modes, 
$  W_\mathrm{Yuk} = Y_{ij}^{(u)} u_i^c q_j H_u + Y_{ij}^{(d)} d_i^c q_j H_d +  
  Y_{ij}^{(l)} l_i e_j^c H_d
$,
where $i,j=1,2,3$ is a family index, then take the form
 \begin{align}
   Y_{ij}^{(u)} &= \left( \begin{array}{ccc}
    a_1 & 0 & a_3 \\
    0 & a_1 & a_3 \\
    a_2 & a_2 & g
   \end{array} \right), &
   Y_{ij}^{(d)} &= \left( \begin{array}{ccc}
    0 & 0 & b_2 \\
    0 & 0 & b_2 \\
    b_5 & b_5 & b_7
   \end{array} \right), &
   Y_{ij}^{(l)} &= \left( \begin{array}{ccc}
    0 & 0 & b_1 \\
    0 & 0 & b_1 \\
    b_3 & b_3 & b_4
   \end{array} \right).
   \nonumber
 \end{align}
Here $g$ denotes the $\SU6$ gauge coupling and is responsible for the large top mass. 
The $a_i$ and $b_i$ represent VEVs of distinct combinations of singlets.
This example illustrates how the mainly unsuccessfull mass relations of standard $\SU5$ GUTs are avoided in our model. It is due to the fact that
the third quark-lepton generation forms two distinct split bulk matter families which
consequently have distinct Yukawa couplings.
\section{Conclusions}
We have constructed a 6D orbifold GUT from the heterotic string. Two of its four fixed points
realize local $\SU5$ GUTs with localized quark-lepton
families. The third family follows from two split bulk generations with distinct couplings to the Higgs doublets.
This avoids
the standard $\SU5$ equality of lepton and down-type quark Yukawa couplings.
It is possible to find vacua with partial or full gauge-Higgs
unification, ensuring a large top Yukawa coupling. However, the implementation of a matter parity into the model needs further investigation.
We calculated the local FI-terms at the fixed points which have crucial influence on the dynamics of bulk fields. Open issues concern their r\^ole in
a blow-up to a smooth manifold and the stabilization of the GUT scale. 
Having in mind the goal of grand unification and the large number of similar compactifications~\cite{lnx07-1,lnx07-2}, 
one may hope that a local GUT structure as investigated here can help to identify realistic string vacua.


\begin{theacknowledgments}
	I would like to thank my collaborators W.~Buchm\"uller and C.~L\"udeling.
\end{theacknowledgments}


\bibliographystyle{aipproc}   
\bibliography{proc-pascos07-jschmidt}

\end{document}